\documentstyle[12pt]{article} 
\normalbaselines
\newcommand\bea{\begin{eqnarray}}
\newcommand\eea{\end{eqnarray}}
\begin{document}
\bibliographystyle{unsrt}

\vbox {\vspace{6mm}} 

\begin{center}
{\large \bf WIGNER DISTRIBUTION FUNCTION AND ENTROPY \\[2mm]
OF THE DAMPED HARMONIC OSCILLATOR \\[2mm]
WITHIN THE THEORY OF OPEN QUANTUM SYSTEMS}\\[7mm]
 A. Isar\\
{\it Department of Theoretical Physics, Institute of
Atomic Physics\\ Bucharest-Magurele, POB MG-6, Romania\\
e-mail address: isar@ifa.ro
}\\[5mm]
{(Presented by G. Nemes)}\\[5mm]
\end{center}

\vspace{2mm}

\begin{abstract}
The harmonic oscillator with dissipation is studied
within the framework of the Lindblad theory for open quantum
systems.
By using the Wang-Uhlenbeck method, the Fokker-Planck equation,
obtained from the master equation for
the density operator, is solved for the Wigner distribution function,
subject to either the Gaussian type or the
$\delta$-function type of initial conditions. The obtained Wigner functions
are two-dimensional Gaussians with different widths.
Then a closed expression for the density operator is extracted.
The entropy
of the system is
subsequently calculated
and its temporal behaviour
shows that this quantity relaxes to its equilibrium value.

\end{abstract}

\section{Introduction}

In the last two decades, the problem of dissipation in quantum
mechanics, i.e. the consistent description of open quantum systems,
was investigated by various authors \cite
{h,d,s,de,l}. Because
dissipative
processes imply irreversibility and, therefore, a preferred direction
in time, it is generally thought that quantum dynamical semigroups are
the basic tools to introduce dissipation in quantum mechanics.
In the Markov approximation
the most general form of the generators of such semigroups was given by
Lindblad \cite
{l1}. This formalism has been studied for the case
of damped
harmonic oscillators \cite{l2,ss,s2} and applied to various physical
phenomena, for instance, the damping of collective modes in deep
inelastic collisions in nuclear physics \cite{i1} and the
interaction of
a two-level atom with the electromagnetic field \cite{s3}.

In the present work, also dealing with the damping of the harmonic
oscillator within
the Lindblad theory for open quantum systems,  we will
explore the physical aspects of the
Fokker-Planck equation which is the $c$-number equivalent equation to
the master equation for the density operator. Generally the master
equation gains considerably in clarity if it is represented in terms
of the Wigner distribution function which satisfies the Fokker-Planck
equation. It is worth mentioning that these master and Fokker-Planck
equations agree in form with the corresponding equations formulated in
quantum optics \cite{a2,gc,sw,k,lz}.

The content of the paper is arranged as follows. In Sec. 2 we review
the derivation of the master equation of the harmonic oscillator.
In Sec. 3 we transform the
master equation into the Fokker-Planck equation by means of the
well-known methods \cite{lo,ha,ga}.  Then the Fokker-Planck equation
for the Wigner
distribution, subject to either the Gaussian type or the
$\delta$-function type of initial conditions, is solved by the
Wang-Uhlenbeck method.
Sec. 4 derives an explicit form of the density operator involved in
the
Lindblad master equation,
formulates the entropy using the
explicit form of the density operator and discusses its temporal
behaviour.
Finally, concluding remarks are given in Sec. 5.

\section{Master equation for the damped harmonic oscillator}

The rigorous formulation for introducing the dissipation into a
quantum mechanical system is that of quantum dynamical semigroups
\cite{d,s,l1}. According to the axiomatic theory of Lindblad
\cite{l1}, the
usual von Neumann-Liouville equation ruling the time evolution of
closed quantum systems is replaced in the case of open systems by the
following equation for the density operator $\rho$:
\bea{d\Phi_{t}(\rho)\over dt}=L(\Phi_{t}(\rho)).\eea
Here, $\Phi_{t}$ denotes the dynamical semigroup describing the
irreversible time evolution of the open system in the Schr\"odinger
representation and $L$ the infinitesimal generator of the dynamical
semigroup $\Phi_t.$ Using the structural theorem of Lindblad \cite{l1}
which
gives the most general form of the bounded, completely dissipative
Liouville operator $L$, we obtain the explicit form of the most
general time-homogeneous quantum mechanical Markovian master equation:
\bea{d\rho(t)\over dt}=L(\rho(t))=-{i\over\hbar}[H,\rho(t)]+{1\over 2
\hbar}\sum_{j}([V_{j}\rho(t),V_{j}^+]+[V_{j},\rho(t)V_{j}^+]).
\eea
Here $H$ is the Hamiltonian of the system and the operators $V_{j}$ and
$V_{j}^+$ are bounded operators on the Hilbert space of the
Hamiltonian.

We should like to mention that the Markovian master equations found in
the literature are of this form after some rearrangement of terms,
even for unbounded Liouville operators. In this connection we assume
that the general form of the master equation given by (2) is also
valid for unbounded Liouville operators.

In this paper we impose a simple condition to the operators $H,V_{j},V_{j}^+$
that they are functions of the basic observables $\hat q$ and $\hat p$ of the
one-dimensional quantum mechanical system (with $[\hat q,\hat p]=i\hbar$) of
such kind
that the obtained model is exactly solvable. A precise version for this last
condition is that linear spaces spanned by first degree (respectively second
degree) noncommutative polynomials in $\hat q$ and $\hat p $ are invariant to
the action
of the completely dissipative mapping $L$. This condition implies \cite{l2}
that $V_
{j}$ are at most first degree polynomials in $\hat q$ and $\hat p $ and $H$ is
at most a
second degree polynomial in $\hat q$ and $\hat p $. Then the harmonic
oscillator
Hamiltonian $H$ is chosen of the form
\bea   H=H_{0}+{\mu \over 2}(\hat q\hat p+\hat p\hat q),~~~H_{0}={1\over 2m}
\hat
p^2
+{m\omega^2\over 2}\hat q^2.     \eea
With these choices the Markovian master equation can be written \cite{ss}:
\bea   {d\rho \over dt}=-{i\over \hbar}[H_{0},\rho]-{i\over 2\hbar}(\lambda +
\mu)
[\hat q,\rho \hat p+\hat p\rho]+{i\over 2\hbar}(\lambda -\mu)[\hat p,\rho
\hat q+\hat q\rho]  \nonumber\\
  -{D_{pp}\over {\hbar}^2}[\hat q,[\hat q,\rho]]-{D_{qq}\over {\hbar}^2}
[\hat p,[\hat p,\rho]]+{D_{pq}\over {\hbar}^2}([\hat q,[\hat p,\rho]]+[\hat p,
[\hat q,\rho]]),      \eea
where $D_{pp},D_{qq}$ and $D_{pq}$ are the diffusion coefficients and $\lambda$
the friction constant. They satisfy the following fundamental constraints
\cite{ss}:
\bea   {\rm i})~D_{pp}>0,~~{\rm ii})~D_{qq}>0,~~{\rm iii})~D_{pp}D_{qq}-
D_{pq}^2
\ge
{{\lambda}^2{\hbar}^2\over 4}.    \eea
In the
particular case when the asymptotic state is a Gibbs state
\bea   \rho_G(\infty)=e^{-{H_0\over kT}}/{\rm Tr}e^{-{H_0\over kT}},    \eea
these coefficients reduce to
\bea   D_{pp}={\lambda+\mu\over 2}\hbar m\omega\coth{\hbar\omega\over 2kT},
{}~~D_{qq}={\lambda-\mu\over 2}{\hbar\over m\omega}\coth{\hbar\omega\over 2kT},
{}~~D_{pq}=0,    \eea
where $T$ is the temperature of the thermal bath.

\section{Wigner distribution function}

One useful way to study the consequences of the master equation (4)
for the density operator of the one-dimensional damped harmonic
oscillator is to transform it into more familiar forms, such as the
equations for the $c$-number quasiprobability distributions Glauber
$P$, antinormal ordering $Q$ and Wigner $W$ associated with the
density operator \cite{i2}. In this case the resulting differential
equations of the Fokker-Planck type for the distribution functions can
be solved by standard methods \cite{lo,ga,r} employed in quantum
optics and
observables directly calculated as correlations of these distribution
functions.

The Fokker-Planck equation, obtained from the master equation and
satisfied by the Wigner distribution function $W(x_1,x_2,t)$
of real variables $x_1, x_2$ corresponding to the operators $\hat q,
\hat p$
\bea x_1=\sqrt{m\omega\over 2\hbar}q, ~~x_2={1\over\sqrt{2\hbar m\omega}}p,
\eea
has the form \cite{i2}:
\bea{\partial W\over\partial t}=\sum_{i,j=1,2}A_{ij}{\partial\over
\partial x_i}(x_jW)+{1\over 2}\sum_{i,j=1,2}Q^W_{ij}{\partial^2\over
\partial x_i\partial x_j}W,  \eea
where
\bea A=\left(\matrix{\lambda-\mu&-\omega\cr
\omega&\lambda+\mu\cr}\right),~~~
Q^W={1\over\hbar}\left(\matrix{m\omega D_{qq}&D_{pq}\cr
D_{pq}&D_{pp}/m\omega\cr}\right).  \eea
Since the drift coefficients are linear in the variables $x_1$ and
$x_2$ and the diffusion coefficients are constant with respect to
$x_1$ and $x_2,$ Eq. (9) describes an Ornstein-Uhlenbeck process
\cite{u,wu}. Following the method developed by Wang and Uhlenbeck
\cite{wu}, we
shall solve this Fokker-Planck equation, subject to either the
wave-packet type or the $\delta$-function type of initial conditions.

1) When the Fokker-Planck equation is subject to a Gaussian
(wave-packet) type of the initial condition  ($x_{10}$ and $x_{20}$ are the
initial values of $x_1$ and $x_2$ at $t=0,$ respectively)
\bea W_w(x_1,x_2,0)={1\over\pi\hbar}\exp\{-2[(x_1-x_{10})^2+(x_2-x_{20})^2]\},
  \eea
the solution is found to be

\bea W_w(x_1,x_2,t)={\Omega\over \pi\hbar\omega\sqrt{\vert B_w\vert}}\exp\{-
{1\over B_w}
[\phi_w(x_1-\bar x_1)^2+\psi_w(x_2-\bar x_2)^2+\chi_w(x_1-\bar x_1)
(x_2-\bar x_2)]\},  \eea
where
\bea   B_w=g_1g_2-{1\over 4}g_3^2,~~g_1=g_2^*={\mu a\over\omega}e^{2\Lambda t}+
{d_1\over \Lambda}
(e^
{2\Lambda t}-1),~~g_3=2[e^{-2\lambda t}+{d_2\over\lambda}(1-e^{-2\lambda t})],
    \eea
\bea\phi_w=g_1a^{*2}+g_2a^2-g_3,~\psi_w=g_1+g_2-g_3,~\chi_w=2(g_1a^*+g_
2a)-g_3(a+a^*).  \eea
We have put $a=(\mu-i\Omega)/\omega,
\Lambda=-\lambda-i\Omega$ and
$d_1=(a^2m\omega D_{qq}+2aD_{pq}+D_{pp}/m\omega)/\hbar,$
$d_2=(m\omega D_{qq}+2\mu D_{pq}/\omega+D_{pp}/m\omega)/\hbar$
and $\Omega^2=\omega^2-\mu^2.$
The functions $\bar x_1$ and $\bar x_2$, which are also oscillating
functions, are given by
\bea\bar x_1=e^{-\lambda t}[x_{10}(\cos\Omega t+{\mu\over\Omega}\sin
\Omega t)+x_{20}{\omega\over\Omega}\sin\Omega t],  \eea
\bea\bar x_2=e^{-\lambda t}[x_{20}(\cos\Omega t-{\mu\over\Omega}\sin
\Omega t)-x_{10}{\omega\over\Omega}\sin\Omega t].  \eea

2) If the Fokker-Planck equation (9) is subject to the
$\delta$-function type of initial condition, the Wigner distribution
function is given by
\bea W(x_1,x_2,t)={\Omega\over \pi\hbar\omega\sqrt{\vert B\vert}}\exp\{-
{1\over B}
[\phi_d(x_1-\bar x_1)^2+\psi_d(x_2-\bar x_2)^2+\chi_d(x_1-\bar x_1)
(x_2-\bar x_2)]\},  \eea
where
\bea B=f_1f_2-f_3^2,~~f_1=f_2^*={d_1\over \Lambda}(e^{2\Lambda t}-1),~~
f_3={d_2\over \lambda}(1-e^{-2\lambda t}),  \eea
\bea\phi_d=f_1a^{*2}+f_2a^2-2f_3,~~\psi_d=f_1+f_2-2f_3,~~\chi_d=2[f_1a^*+f_
2a-f_3(a+a^*)].  \eea
So, the Wigner functions are 2-dimensional Gaussian distributions with the
average
values $\bar x_1$ and $\bar x_2$ and different widths.

When time $t\to\infty,$ $\bar x_1$ and $\bar x_2$ vanish and
we obtain the steady state solution:
\bea W(x_1,x_2)={1\over 2\pi\sqrt{{\rm det}\sigma^W(\infty)}}\exp[-{1
\over 2}\sum_{i,j=1,2}(\sigma^W)^{-1}_{ij}(\infty)x_ix_j].
  \eea
The stationary covariance matrix $\sigma^W(\infty)$ can be determined
from the algebraic equation
\bea A\sigma^W(\infty)+\sigma^W(\infty)A^{\rm T}=Q^W.  \eea

\section{Entropy and effective temperature}

Entropy is a quantity which may be visualized physically as a measure
of the lack of knowledge of the system.
When we denote by $ \rho(t)$ the density operator in the Schr\"odinger
picture for the harmonic oscillator, the entropy  $S(t) $ is given by
\bea   S(t)= -k{\rm Tr}(\rho\ln\rho).\eea
For calculating the entropy we shall
compute straightway
the expectation value of the logarithmic operator
$<\ln\rho>={\rm Tr}(\rho\ln\rho).$
Accordingly, the problem amounts to derive
the explicit form of the density operator for the damped harmonic oscillator.

To get the explicit expression for the density operator, we use the relation
$\rho=2\pi\hbar {\bf N}\{W_s(q,p)\},$ where $W_s$ is the Wigner
distribution
function in the form of standard rule of association and ${\bf N}$ is the
normal ordering operator \cite{lo,w} which acting on the function $W_s(q,p)$
moves all $p$ to the right of the $q.$ By the standard
rule of association is meant the correspondence $p^mq^n\to \hat q^n\hat p^m$
between functions of two classical variables $(q,p)$ and functions of two
quantum mechanical canonical operators $(\hat q,\hat p).$ The calculation of
the density operator is then reduced to a problem of transformation of the
Wigner distribution function by the ${\bf N}$ operator, provided that $W_s$
is
known. A special care is necessary for the ${\bf N}$ operation when the
Wigner
function is in the exponential form of a second order polynomial of $q$ and
$p.$ The Wigner distribution function
previously obtained corresponds however to the form of the
Weyl rule of association \cite{hw}.
 The solution (12)
of the Fokker-Planck equation (9), subject to the wave-packet type of initial
condition (11) can be written in terms of the coordinate and momentum as:

\bea   W(q,p,t)={1\over 2\pi\sqrt{\delta}}\exp\{-{1\over \delta}[\phi
(q-<\hat q>)^2+\psi(p-<\hat p>)^2-2\chi(q-<\hat q>)(p-<\hat p>)]\},
    \eea
where
\bea <\hat q>= \sqrt{{2\hbar\over m\omega}}\bar x_1,
{}~~<\hat p>= \sqrt{2\hbar m\omega}\bar x_2,\eea

\bea  \phi\equiv\sigma_{pp}=<\hat q ^2>-<\hat q >^2
=-{\hbar\omega^2\over 4\Omega^2}{1\over m\omega}\psi_w,
,\eea

\bea  \psi\equiv\sigma_{qq}=<\hat p ^2>-<\hat p >^2=
-{\hbar\omega^2\over 4\Omega^2}m\omega\phi_w,
\eea
\bea  \chi\equiv\sigma_{pq}(t)={1\over 2}<\hat q \hat p +\hat p \hat q >-
<\hat q ><\hat p >
={\hbar\omega^2\over 8\Omega^2}\chi_w,~~\delta=\phi\psi-\chi^2
\eea
and $<\hat A>={\rm Tr}(\rho \hat A)$ denotes the expectation value of an
operator $\hat A.$
The Wigner distribution function (23) can be transformed into the form of
standard rule of association \cite{clm} by
\bea   W_s(q,p)=\exp({1\over 2}i\hbar{\partial^2\over\partial p
\partial q})W(q,p).
\eea
Upon performing the operation on the right-hand side, we get the Wigner
distribution function $W_s,$ which has the same form as the original $W$
multiplied by $\hbar$ but with $\chi-i\hbar/2$ in place of
$\chi.$
The normal ordering operation of the Wigner function $W_s$ in
Gaussian form can be carried out by applying McCoy theorem \cite{w,m,j}.
The explicit form of the density operator is the following:

  \bea  \rho={\hbar\over \sqrt{\xi}}\exp \bigl[{1\over 2}\ln{\xi\over \xi-i
\hbar
\chi'}
-{1\over 2\hbar\sqrt{\xi-i\hbar\chi'+{1\over 4}\hbar^2}}\cosh^{-1}(1+{\hbar^2
\over 2(\xi-i\hbar\chi')})\nonumber\\
  \times\{\phi(\hat q-<\hat q>)^2+\psi(\hat p-<\hat p>)^2-(\chi'+i{\hbar\over
2})
[2(\hat q-<\hat q>)(\hat p-<\hat p>)-i\hbar]\}\bigr],
      \eea
where
\bea  \xi=\phi\psi-\chi'^2,
{}~~\chi'=\chi-i{\hbar\over 2}.\eea
The density operator (29) is in a Gaussian form, as was expected from the
initial form of the Wigner distribution function. While the density operator is
expressed in terms of operators $\hat q$ and $\hat p,$ the Wigner distribution
is a function of real variables $q$ and $p.$ When time $t$ goes to infinity,
the density operator approaches to
\bea  \rho(\infty)={\hbar\over \sqrt{\sigma-{\hbar^2\over 4}}}\exp[-{1\over 2
\hbar
\sqrt{
\sigma}}\ln{2\sqrt{\sigma}+\hbar\over 2\sqrt{\sigma}-\hbar}
[\sigma_{pp}(\infty)\hat q^2+\sigma_{qq}(\infty)\hat p^2-\sigma_{pq}(\infty)
(\hat q\hat p+\hat p\hat q)],
    \eea
where $\sigma=\sigma_{pp}(\infty)\sigma_{qq}(\infty)-\sigma^2_{pq}(\infty)$
and [8]:
\bea   \sigma_{pp}(\infty)={1\over 2\lambda(\lambda^2+\omega^2-\mu^2)}((m
\omega)^2
\omega^2D_{qq}+(2\lambda(\lambda-\mu)+\omega^2)D_{pp}-2m\omega^2(\lambda-
\mu)D_{pq}),    \eea
\bea   \sigma_{qq}(\infty)={1\over 2(m\omega)^2\lambda(\lambda^2+\omega^2-\mu^
2)}
((m\omega)^2(2\lambda(\lambda+\mu)+\omega^2)D_{qq}
  +\omega^2D_{pp}+2m\omega^2(\lambda+\mu)D_{pq}),\eea

\bea   \sigma_{pq}(\infty)={1\over 2m\lambda(\lambda^2+\omega^2-\mu^2)}(-(
\lambda+
\mu)(m\omega)^2D_{qq}+(\lambda-\mu)D_{pp}+2m(\lambda^2-\mu^2)D_{pq}).\eea
In the particular case (7)

\bea  \sigma_{qq}(\infty)={\hbar\over 2m\omega}\coth{\hbar\omega\over 2kT},~
\sigma_{pp}(\infty)={\hbar m\omega\over 2}\coth{\hbar\omega\over 2kT},~
 \sigma_{pq}(\infty)=0    \eea
and the asymptotic state is a Gibbs state (6):

\bea  \rho_G(\infty)=2\sinh{\hbar\omega\over 2kT}\exp[-{1\over kT}
({1\over 2m}\hat p^2+{m\omega^2\over 2}\hat q^2)].\eea
Because of the presence of the exponential form in the density operator, the
construction of the logarithmic density is straightforward.
In view of the relations (25-27), the expectation value of the logarithmic
density becomes
\bea  <\ln\rho>=\ln\hbar-{1\over 2}\ln(\delta-{\hbar^2\over 4})-{\sqrt{\delta}
\over
\hbar}\ln{2\sqrt{\sigma}+\hbar\over 2\sqrt{\sigma}-\hbar}.\eea
By putting $\hbar\nu=\sqrt{\delta}-\hbar/2,$
we finally get the entropy in a closed form:
\bea  S(t)=k[(\nu+1)\ln(\nu+1)-\nu\ln\nu].    \eea
It is worth noting that the entropy depends only upon the variance of the
Wigner distribution.
When time $t\to\infty,$ the function $\nu$ goes to $s=\omega(d_2^2/
\lambda^2-\vert d_1\vert^2/(\lambda^2+\Omega^2))^{1/2}/2\Omega-1/2$
and the entropy relaxes to its equilibrium value $S(\infty)=k[(s+1)\ln(s+1)-
s\ln s].$
It should also be noted that the expression (38) has the same form as the
entropy of a system of harmonic oscillators in thermal equilibrium. In the
later case $\nu$ represents, of course, the average of the number
operator \cite{a3}. While the formal expression (38) for the entropy has a
well-known
appearance, the form of the function $\nu$ displays clearly a
specific
feature of the present entropy.
We see that the time
dependence of the entropy is represented by the damping factor $\exp(-2\lambda
t)$ and also by the oscillating function $\sin^2(\Omega t).$
The entropy relaxes to its equilibrium value $S(\infty).$

\section{Concluding remarks}

Recently we assist to a revival of interest in quantum Brownian motion
as a paradigm of quantum open systems. There are many motivations. The
possibility of preparing systems in macroscopic quantum states led to
the problems of dissipation in tunneling and of loss of quantum
coherence (decoherence). These problems are intimately related to the
issue of quantum-to-classical transition. All of them point the
necessity of a better understanding of open quantum systems and all
requires the extension of the model of quantum Brownian motion. The
Lindblad theory provides a selfconsistent treatment of damping as a
possible extension of quantum mechanics to open systems. In the
present paper we have studied the one-dimensional harmonic oscillator
with dissipation within the framework of this theory. From the master
equation of the damped quantum oscillator we have
derived the corresponding Fokker-Planck equation in the Wigner $W$
representation. The obtained equation describes an Ornstein-Uhlenbeck
process. By using the Wang-Uhlenbeck method we have solved this
equation for the Wigner function, subject to either the Gaussian type
or the $\delta$-function type of initial conditions and showed that
the Wigner functions are two-dimensional Gaussians with different
widths. Then we have obtained the density operator.
 The density operator
in a Gaussian form is a function of $\hat q,\hat p$ in addition to several
time dependent
factors. The explicit form of the density operator has been subsequently used
to calculate the entropy. It relaxes
to its equilibrium value.

\end{document}